\documentclass[journal]{IEEEtran}
\usepackage{eurosym}
\usepackage{amssymb}
\usepackage{amsmath}
\usepackage{textcomp}
\usepackage[caption=true]{caption}
\usepackage{subcaption}
\usepackage{epsfig}
\usepackage{amsthm}
\usepackage{mathtools}
\usepackage{multirow}
\usepackage{float}
\usepackage{color}
\usepackage{cite}
\usepackage{graphicx}
\usepackage{epstopdf}
\usepackage{etoolbox}

\newtheorem{algorithm}{Algorithm}

\AtBeginEnvironment{align}{\setcounter{subeqn}{0}}
\newcounter{subeqn}

\begin{document}

\title{Intelligent Pinning Based Cooperative Secondary Control of
Distributed Generators for Microgrid in Islanding Operation Mode}
\author{M. Talebi$^\dagger$\thanks{$\dagger$ M. Talebi is with the ECE
department at the University of Central Florida (UCF), Orlando, FL. E-mail: 
\texttt{\small morteza@knights.ucf.edu}. }, S.~Manaffam$^{\ddagger}$%
\thanks{$^\ddagger$ S. Manaffam is with the NanoScience Technology Center,
UCF, Orlando, FL. E-mail: \texttt{\small saeedmanaffam@knights.ucf.edu}.},
A. K. Jain$^{\dagger\dagger}$\thanks{$^{\dagger\dagger}$ A. K. Jain is
with the Client Computing Group at Intel Corporation, Portland, OR. E-mail: 
\texttt{amit\_k\_jain@ieee.org}.}, and~A.~Behal$^{\ddagger\ddagger}$%
\thanks{
$^{\ddagger\ddagger}$ A. Behal (corresponding author)is with 
ECE and NSTC, UCF, Orlando, FL. E-mail: \texttt%
{\small abehal@ucf.edu.}}}
\maketitle

\begin{abstract}
Motivated by the fact that the location(s) and structural properties of the
pinning node(s) affect the algebraic connectivity of a network with respect
to the reference value and thereby, its dynamic performance, this paper
studies the application of intelligent single and multiple pinning of
distributed cooperative secondary control of distributed generators (DGs) in
islanded microgrid operation. It is shown that the intelligent selection of
a pinning set based on the degree of connectivity and distance of leader
DG(s) from the rest of the network improves the transient performance for
microgrid voltage and frequency regulation. The efficacy of the distributed
control strategy based on the proposed algorithms is illustrated via
numerical results simulating typical scenarios for a variety of microgrid
configurations.
\end{abstract}


\section{Introduction}

\IEEEPARstart{V}{oltage} and frequency regulation of microgrid are essential
in both grid connected and islanded modes \cite{Mehrizi,Hirar,Davoudi}. In
grid connected mode, frequency is dictated by the main grid while the
voltage within the microgrid can be regulated based on its reactive power
generation and consumption. When the microgrid disconnects from the main
grid in response to, say, upstream disturbance or voltage fluctuation and
goes to islanding mode, both voltage and frequency at all locations in the
microgrid have to be regulated to nominal values. Typical microgrid control
hierarchy includes (1) primary control for real and reactive power sharing
between distributed generators (DGs) \cite{Brab,Divan,Green,Marwali}; and
(2) secondary control to maintain load voltage and frequency close to
nominal values via respective references communicated to each DG \cite%
{Davoudi}\!\!\cite{Olivar} from a
controller. Using centralized control requires complex and expensive
communication infrastructure and is subject to failure in the central
controller and communication links. To overcome these limitations,
distributed cooperative control \cite{Ren,Olfati,Xin} can be utilized which employs a sparse network. The
application of cooperative control for DG operation in power systems in grid
connected mode has been studied in \cite{Xin,Simp,Lu,Kim}
among others. The technique of providing the reference values to only a
fraction of the DGs in the network, known as \textit{pinning}, has been
studied in \cite{Su}\!\!\cite{Li}. In \cite{Liu}, 
\textit{pinning control} in microgrid islanding mode using energy storage
system as master unit was recently studied. In \cite{Bidram}, Bidram \textit{%
et al.} employed multi-agent distributed cooperative secondary voltage
control of the microgrid.

In the aforementioned studies, the selection of the pinned (or leader) DG(s)
has been assumed to be arbitrary. However, as has been shown in \cite%
{Manaffam13a,DeLellis13,Chen07}, the performance and robustness of the
network is directly related to the choice of the \textit{pinning set, } 
\textit{i.e.,} the set of pinned or leader DG(s). Recent results obtained by
the authors in \cite{Saeed} obtain tight upper and lower bounds on the
algebraic connectivity of the network to the reference signal\footnote{%
Algebraic connectivity to the reference signal \cite{DeLellis13} is a
measure of the availability of the reference information in the overall
network.}. By taking advantage of these novel results, the work in this
paper proffers multiple novel contributions. We first formulate the problem
of single and multiple pinning of multi-agent distributed cooperative
control in microgrids. Next, the effect of proper selection of pinning DGs
is discussed. Then, we propose two implementable pinning node(s) selection
algorithms based on degree and distance of the candidate leader(s) from the
rest of the microgrid. Several cases of power system topologies and
communication networks as well as different scenarios are numerically
simulated to show that intelligent selection of the pinning node(s) results
in better transient performance of the DGs' terminal profiles.

The remainder of the paper is organized as follows. Section II reviews the
microgrid primary and secondary control schemes for regulating voltage and
frequency. The intelligent single and multiple pinning problems are
formulated in Section III followed in Section IV by a presentation of the
algorithms for solving these problems. Section V illustrates the performance
of the proposed algorithms via extensive numerical simulation studies while
section VI concludes the paper.

\section{Microgrid System Model}

In this section, we briefly describe the basic system model consisting of
the inverter based DG, the primary controller for the DG, and the
distributed cooperative secondary controller. For more details, the reader
is referred to the development in \cite{Bidram}\!\!\cite{BidramCoop} and the
references therein.

\subsection{DG\ Model and Primary Control}

The inverter based DG model consists of a three legged inverter bridge
connected to a DC voltage source. The DC bus dynamics and switching process
of the inverter can be neglected due to the assumptions of an ideal DC
source and realization of high switching frequency of the bridge,
respectively. The output frequency and voltage magnitude of each DG are set
in accordance with the droop controllers \cite{Divan}. The output voltage
magnitude and frequency control of the inverter itself is typically
implemented with internal current controllers in the standard $d-q$ rotating
reference frame with the d-axis aligned to the output voltage vector, 
\textit{e.g.,} see \cite{Poga}. From the droop control point of view, the
inverter can be assumed to be ideal in the sense that its output voltage
magnitude and frequency are regulated to the desired values of $%
V_{i,mag}^{\star }$ and $\omega _{i}^{\star }$, respectively, within the
time frame of interest. Thus, 
\begin{equation}
\begin{array}{l}
V_{od_{i}}=V_{i,mag}^{\star }=V_{n_{i}}-n_{Q_{i}}Q_{i};V_{oq_{i}}^{\star }=0
\\ 
\omega _{i}=\omega _{i}^{\star }=\omega _{n_{i}}-m_{P_{i}}P_{i}%
\end{array}%
.  \label{ndg}
\end{equation}%
where $\omega _{i}^{\star }$ and $V_{i,mag}^{\star }$ are the desired
angular frequency and voltage amplitude of the $i^{th}$ DG, respectively; $%
P_{i}$ and $Q_{i}$ are the active and reactive power outputs of the $i^{th}$
DG; $\omega _{n_{i}}$ and $V_{n_{i}}$ are reference angular frequency and
voltage set points determined by the secondary control, respectively; and $%
m_{P_{i}}$ and $n_{Q_{i}}$ are droop coefficients for real and reactive
power. Here, $P_{i}$ and $Q_{i}$ are generated by passing the instantaneous
real and reactive power outputs of the inverter through first order low pass
filter with corner frequency $\omega _{c_{i}}$ to eliminate undesired
harmonics%
\begin{equation}
\begin{array}{c}
\dot{P}_{i}=-\omega _{c_{i}}P_{i}+\omega
_{c_{i}}(v_{od_{i}}i_{od_{i}}+v_{oq_{i}}i_{oq_{i}}) \\ 
\dot{Q}_{i}=-\omega _{c_{i}}Q_{i}+\omega
_{c_{i}}(v_{oq_{i}}i_{od_{i}}-v_{od_{i}}i_{oq_{i}})%
\end{array}%
.  \label{PQ}
\end{equation}

\subsection{Distributed Cooperative Secondary Control}

\subsubsection{Secondary Voltage and Frequency Control}

The goal in secondary voltage control is to determine the primary control
reference voltage $V_{n_{i}}$ inside each DG so as to minimize the deviation
of the concerned DG's output voltage $V_{od_{i}}$ from the constant nominal
reference voltage $V_{ref}$. Mathematically, this is accomplished by the
following set of equations%
\begin{equation}
\begin{array}{rl}
\dot{V}_{n_{i}} & =u_{v_{i}}+n_{Q_{i}}\dot{Q}_{i} \\ 
\Rightarrow V_{n_{i}} & =\int (u_{v_{i}}+n_{Q_{i}}\dot{Q}%
_{i})dt=n_{Q_{i}}Q_{i}+\int (u_{v_{i}})dt%
\end{array}
\label{eq:V_ni_calc}
\end{equation}%
where $u_{v_{i}}$ is an auxiliary control input given by 
\begin{equation}
u_{v_{i}}=-c_{v}\sum_{j=1}^{N}a_{ij}(e_{v_{i}}-e_{v_{j}})-c_{v}g_{i}\zeta
_{i}e_{v_{i}}  \label{eq: v_input}
\end{equation}%
where $e_{v_{i}}\triangleq V_{od_{i}}-V_{ref}$ is the voltage regulation
error for $i^{th}$ DG , $c_{v}>0$ is a control gain, while $N$ denote the
total number of DGs in the network. Here, $\mathbf{A}=[a_{ij}]$ is the
adjacency matrix of the communication network, $a_{ij}=1$ indicates a
connection from $j^{th}$ DG to $i^{th}$ DG and otherwise $a_{ij}=0$; $g_{i}$
is the pinning gain of the $i^{th}$ DG and $\zeta _{i}\in \{0,\,1\}$
indicates which nodes are pinned to the reference value.

\begin{figure*}[t]
\center
\includegraphics[scale=.35]{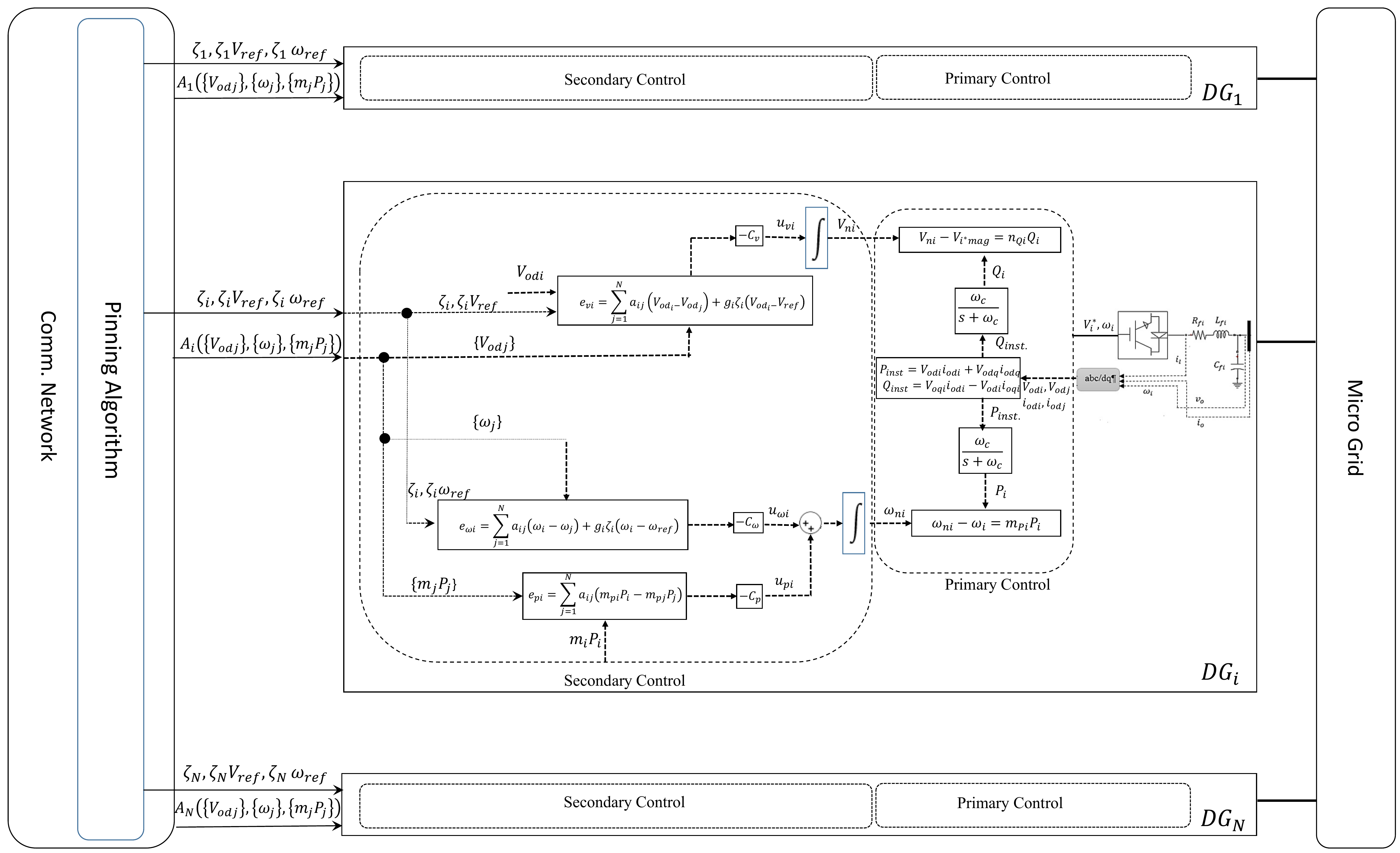}
\caption{Distributed cooperative secondary control in a microgrid ($A_{i}$
denotes $i^{th}$ row of communication network adjacency matrix, $A$). }
\label{fig:distrib-coop-sec-ctrl}
\end{figure*}

For the secondary frequency control, the twin objectives are to ensure that
the frequency of each DG\ reaches the reference frequency $\omega _{ref}$
while at the same time ensuring that active power is shared in proportion to
the rating of the individual DGs. In other words, leader-following consensus
and leaderless consensus problems are solved as follows 
\begin{equation}
\dot{\omega}_{n_{i}}=u_{\omega _{i}}+u_{p_{i}}\Rightarrow \omega
_{n_{i}}=\int (u_{\omega _{i}}+u_{p_{i}})dt  \label{eq: f_input}
\end{equation}%
where $u_{\omega _{i}}$ is the first of two auxiliary control inputs given by%
\begin{equation}
u_{\omega _{i}}=-c_{\omega }\sum_{j=1}^{N}a_{ij}(e_{\omega _{i}}-e_{\omega
_{j}})-c_{\omega }g_{i}\zeta _{i}e_{\omega _{i}},  \notag
\end{equation}%
where $c_{\omega }>0$ is the control gain and $e_{\omega _{i}}\triangleq
\omega _{i}-\omega _{ref}$ is the frequency regulation error of the $i^{th}$
DG with respect to the reference signal. The second auxiliary control term $%
u_{p_{i}}$ (added to ensure that generators share active power in accordance
with their ratings as captured in the droop curve coefficients $m_{P_{i}}$)
is generated as follows 
\begin{equation}
u_{p_{i}}=-c_{p}\sum_{j=1}^{N}a_{ij}(m_{P_{i}}{P}_{i}-m_{P_{j}}{P}_{j})
\label{eq: power_sharing}
\end{equation}%
where $c_{p}>0$ is a control gain. A block schematic of the microgrid with
voltage source inverter (VSI) based DGs alongside the primary and secondary
controllers is shown in Fig. \ref{fig:distrib-coop-sec-ctrl}. 

\subsubsection{Error Dynamics}

After substituting (\ref{eq: v_input}) and (\ref{eq: f_input}) into the time
dynamics of (\ref{ndg}), we have 
\begin{eqnarray}
\dot{\mathbf{e}}_{v} &=&-c_{v}(\mathbf{L}+\mathbf{G\,Z})\mathbf{e}_{v},
\label{eq: v_error} \\
\dot{\mathbf{e}}_{\omega } &=&-c_{\omega }(\mathbf{L}+\mathbf{G\,Z})\mathbf{e%
}_{\omega }  \label{eq: f_error}
\end{eqnarray}%
where ${\mathbf{e}}_{v}\triangleq \lbrack e_{v_{1}},e_{v_{2}},\cdots
,e_{v_{N}}]^{T}$ and ${\mathbf{e}}_{\omega }\triangleq \lbrack e_{\omega
_{1}},e_{\omega _{2}},\cdots ,e_{\omega _{N}}]^{T}$ denote the output
voltage and frequency regulation errors of the network, $\mathbf{G}%
\triangleq \text{diag}([g_{1},\,g_{2},\,\cdots ,\,g_{N}])$ is the network's
pinning gain matrix, $\mathbf{Z}\triangleq \text{diag}([\zeta _{1},\,\zeta
_{2},\,\cdots ,\,\zeta _{N}])$ is the pinning matrix, and $\mathbf{L}%
\triangleq \lbrack l_{ij}]$ is the Laplacian of the adjacency matrix defined
as 
\begin{equation}
l_{ij}\triangleq \left\{ 
\begin{array}{ll}
-a_{ij} & i\neq j, \\ 
\sum_{j=1}^{N}a_{ij} & i=j.%
\end{array}%
\right.   \label{eq: Laplacian}
\end{equation}%
From (\ref{eq: v_error}) and (\ref{eq: f_error}), it is clear that pinning
choices $g_{i}$ and $\zeta _{i}$ are critical to the performance of the
secondary controller.

\section{Pinning Problem}

\subsection{Problem Formulation}

As is well known from linear time-invariant system theory, the performance
and robustness of the systems in (\ref{eq: v_error}) and (\ref{eq: f_error})
are directly dependent on the eigenvalues of $\mathbf{L}+\mathbf{G\,Z}$.
Therefore, choosing which DG(s) to provide the reference values to, \textit{%
i.e.}, $\mathbf{Z}$, has an enormous impact on the performance of the
microgrid. The first problem (Pinning Problem 1) that is addressed in this
paper is that of choosing the location of the nodes to be pinned under the
specification of a given size of the pinning set. In other words, given a
desired number of desired pinning nodes $m$, one needs to determine the $%
\mathbf{Z}$ that maximizes the minimum eigenvalue of the closed-loop system.
The second problem (Pinning Problem 2) addressed in this paper is to find
the minimum number of pinning nodes (and their locations) while guaranteeing
a certain specified convergence rate $\lambda ^{\star }$.

\subsection{Solutions to the Pinning Problems}

It is well known that the optimal solutions for the aforementioned problems
have exponential complexity \cite{Yu}. Thus, finding the optimal solution is
not practical in microgrids with large number of DGs which renders
suboptimal solutions with polynomial time complexity to be of immense
interest \cite{Saeed,Manaffam13a,DeLellis13,Chen07}. To develop any
effective algorithm for either of the pinning problems stated above, we need
to understand how the structural properties (such as distance, number of
connections, \textit{etc.}) of the pinning set in the network affects the
algebraic connectivity, $\phi (\mathbf{Z})\triangleq \lambda _{\min }(%
\mathbf{L}+g\mathbf{Z})$, of the network with respect to the reference. One
possible way to understand these effects is by utilizing lower and upper
bounds on $\phi (\mathbf{Z})$. A set of tight upper and lower bounds on the
algebraic connectivity of the network with respect to the reference have
recently been developed in \cite{Saeed} and are provided in Appendix \ref%
{app: bounds} for the case of $m$ pinned nodes with $1\leq m\leq N$. In what
follows, we discuss the implications of these bounds.

For illustration, consider first the single pinning case where the reference
value is only available in one DG (\textit{i.e.,} $m=1$). Based on the
bounds in Appendix \ref{app: bounds}, it is clear that the upper bound on $%
\phi (\mathbf{Z})$ is a strictly increasing function of the degree of the
pinning node and its pinning gain $g$. Additionally, if the pinning gain, $g$%
, is considered to be very large, the upper bound on algebraic connectivity
of the network with respect to the reference can be shown to be bounded as $%
d_{i}/N$ where $d_{i}$ denotes the out degree of the pinning node. Since $%
d_{i}\leq N-1$, the algebraic connectivity of the network with respect to
the reference cannot exceed $1$ for the single pinning case. Thus, it is
clear to see that the convergence rate of the errors in (\ref{eq: v_error})
and (\ref{eq: f_error}) cannot exceed $c_{v}$ and $c_{\omega }$ for voltage
and frequency, respectively. Since we established that the algebraic
connectivity of the network with respect to the reference is upper bounded
by $d_{i}\leq N-1$, it can be conjectured that pinning a DG with higher
out-degree is more effective than pinning a DG with lower out-degree.
Furthermore, from the given lower bound on $\phi (\mathbf{Z})$ in Appendix %
\ref{app: bounds}, it can be concluded that another topological
characteristic to choose the pinning DG is its \textit{centrality} value
which is a measure of its distance from the farthest DG in the network. This
means that the candidate DG for pinning should have a low distance from the
rest of the network. The general case of multiple pinning (\textit{i.e.,} $%
m>1$) can be understood by considering the set of pinned DGs as a supernode
in the network and by calculating all the topological properties according
to this modified definition.

Before we proceed to state the algorithms arising from these bounds, let us
state the following nomenclature

\begin{eqnarray*}
\text{path-length}(i,j) &=&\text{length of shortest path from DG $i$ to } \\
&&\text{DG $j$,} \\
\text{path}(\mathcal{P},\,\mathcal{I}) &=&\sum_{j\in \mathcal{I}}\min_{i\in 
\mathcal{P}}(\text{path-length}(i,j)) \\
\text{deg}(\mathcal{P}) &=&\sum_{j\in \mathcal{P}}\sum_{i\in \mathcal{I}%
}a_{ij}
\end{eqnarray*}%
where $\mathcal{N}=\{1,\,\cdots ,\,N\}$ is the set of all DGs, $\mathcal{P}$
denotes the pinning set, $\mathcal{I}=\mathcal{N}\setminus \mathcal{P}$
where $\setminus $ denotes minus operation for sets, while d$\text{eg}(%
\mathcal{P})$ is the number of links directed from the pinning DGs to the
rest of the network.

\subsubsection{Suboptimal Algorithm for Pinning Problem 1}

This algorithm will choose $m$ pinning DGs to maximize the minimum
eigenvalue of the closed-loop system. Since the upper bound on the minimum
eigenvalue of the pinned network is an increasing function of the degree of
the pinning DG, the DGs are sorted by decreasing degree; furthermore, from (%
\ref{eq: lower_multiple}), we know that the minimum path length between the
pinning set and the rest of the network should be minimized. Hence, the
added pinning DG to the pinning set should maximize the combined measure,
which is the minimum path length between the candidate pinning DG and the
rest of the network subtracted from the out-degree of pinning. This
procedure should continue until there $m$ DGs in the pinning set. The
pseudo-code of this algorithm can be given as

\begin{algorithm}
~

\begin{enumerate}
\item {$\mathcal{P}= \emptyset$ and $\mathcal{I} = \mathcal{N}$, }

\item {\ while $|\mathcal{P}|<m$ do }

\begin{itemize}
\item {$i^{\star }=\underset{\forall i\in \mathcal{I}}{\text{argmax}}\Big($d$%
\text{eg}(\mathcal{P}\cup \{i\})-\text{path}(\mathcal{P}\cup \{i\},\mathcal{I%
}\setminus \{i\})\Big)$ }

\item {\ $\mathcal{P}=\mathcal{P}\cup \{i^\star\}$, $\mathcal{I}=\mathcal{I}%
\setminus \{i^\star\}$. }
\end{itemize}
\end{enumerate}
\end{algorithm}

\begin{figure}[t]
\centering
\begin{subfigure}{.45\textwidth}
		\centering
		\includegraphics[width=2.9in]{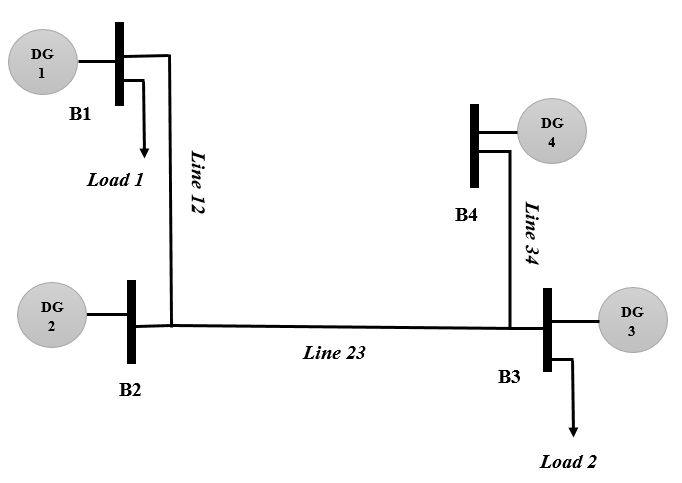}
		\caption{Single line diagram of the system configuration.}
		\label{fig: 4bus}
	\end{subfigure} \hspace{0.05\textwidth} 
\begin{subfigure}{.35\textwidth}
		\centering
		\includegraphics[width=1.25in]{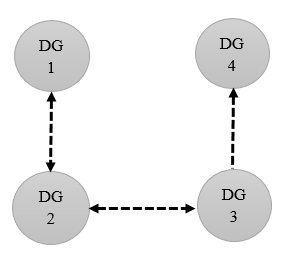}
		\caption{communication network}\label{fig: 4busCom}
	\end{subfigure}
\caption{4-bus power system configuration (dashed arrows represent
information flow).}\vspace{-.5cm}
\label{fig:4buspinning}
\end{figure}

\subsubsection{Suboptimal Algorithm for Pinning Problem 2}

In the second problem, let $\mu ^{\star }$ be the desired algebraic
connectivity of the network to the reference. Since the algebraic
connectivity of the network to the reference cannot exceed the maximum
out-degree of the pinning set, the minimum number of pinning DGs to achieve
a target convergence rate is given as $\mu ^{\star }\leq {\text{deg}(%
\mathcal{P})/}\left( N-1\right) $. Consequently, the smallest number of
pinning DGs, $m$, to achieve the desired convergence rate should be chosen
such that summation of the $m$ highest degree DGs should exceed $(N-1)\mu
^{\star }$. Algorithm 1 is started with this $m$ and arrives at a pinning
set $\mathcal{P}$. If the condition $\phi (\mathbf{Z})\geq \mu ^{\star }$ is
satisfied, the algorithm stops, otherwise, one more pinning DG is added
using Algorithm 1 until the desired convergence rate is achieved. The
pseudo-code of this algorithm can be expressed as

\begin{algorithm}
~

\begin{enumerate}
\item {Sort the degree of the DGs such that 
\begin{equation*}
d_{i_1}\ge \cdots \ge d_{i_N},
\end{equation*}%
}

\item {Set $m$ to be smallest integer such that 
\begin{equation*}
\sum_{j=1}^{m} d_{i_j}\ge (N-1)\mu^\star,
\end{equation*}%
}

\item {$\mathcal{P}= \emptyset$ and $\mathcal{I} = \mathcal{N}$, }

\item {\ while $|\mathcal{P}|<m$ do }\label{step1}

\begin{enumerate}
\item {$i^\star = \underset{\forall i \in\mathcal{I}}{\text{argmax}}\Big(%
\text{Deg}(\mathcal{P}\cup \{i\})-\text{path}(\mathcal{P}\cup\{i\},\mathcal{I%
}\setminus\{i\})\Big)$ }

\item {\ $\mathcal{P}=\mathcal{P}\cup \{i^\star\}$, $\mathcal{I}=\mathcal{I}
\setminus \{i^\star\}$. }
\end{enumerate}

\item {if $\mathbf{L}+g\mathbf{Z}-\mu ^{\star }\mathbf{I}_{N}\succeq \mathbf{%
0}$, then stop; \newline
else: \quad set $m=m+1$ and go to \ref{step1}.}
\end{enumerate}
\end{algorithm}

\section{Numerical Results\label{sec:selforg}}


\subsection{Grid and Controller Parameters}

\begin{figure}[t]
\centering
\begin{subfigure}{.55\textwidth}
		\centering
		\includegraphics[width=2.350in]{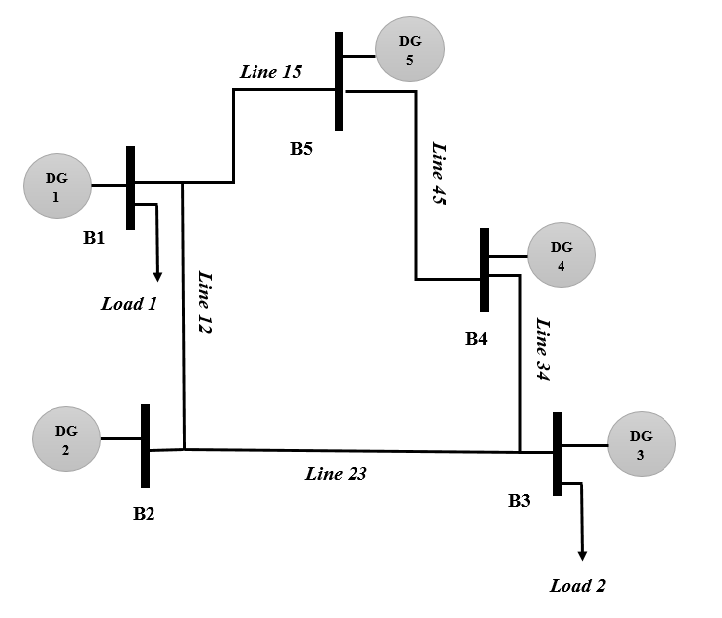}
		\caption{Single line diagram of the system configuration.}
		\label{fig:5}
	\end{subfigure}
\begin{subfigure}{.4\textwidth}
		\centering
		\includegraphics[width=1.2in]{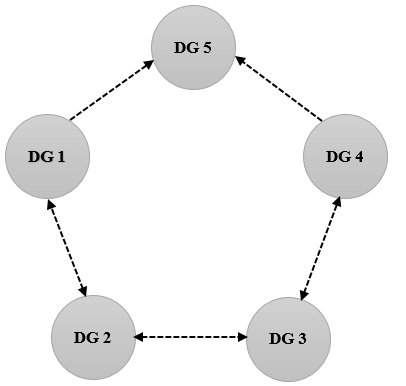}
		\caption{Communication network .}
		\label{fig:5b}
	\end{subfigure}
\caption{5-bus power system configuration. }
\label{fig:5busRing}
\end{figure}
\begin{figure}[tbp]
\centering
\includegraphics[width=8.5cm,height=3.8cm]{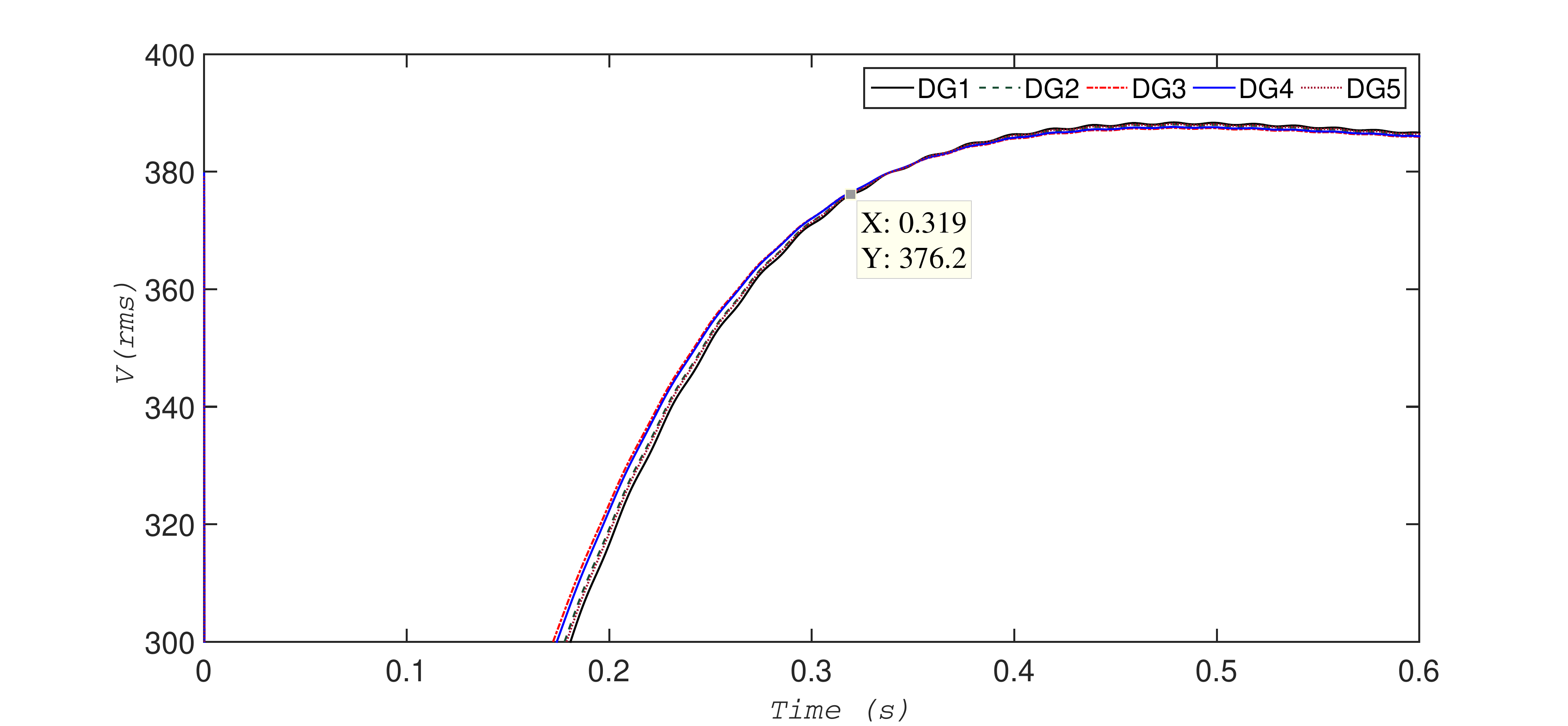} %
\includegraphics[width=8.5cm,height=3.8cm]{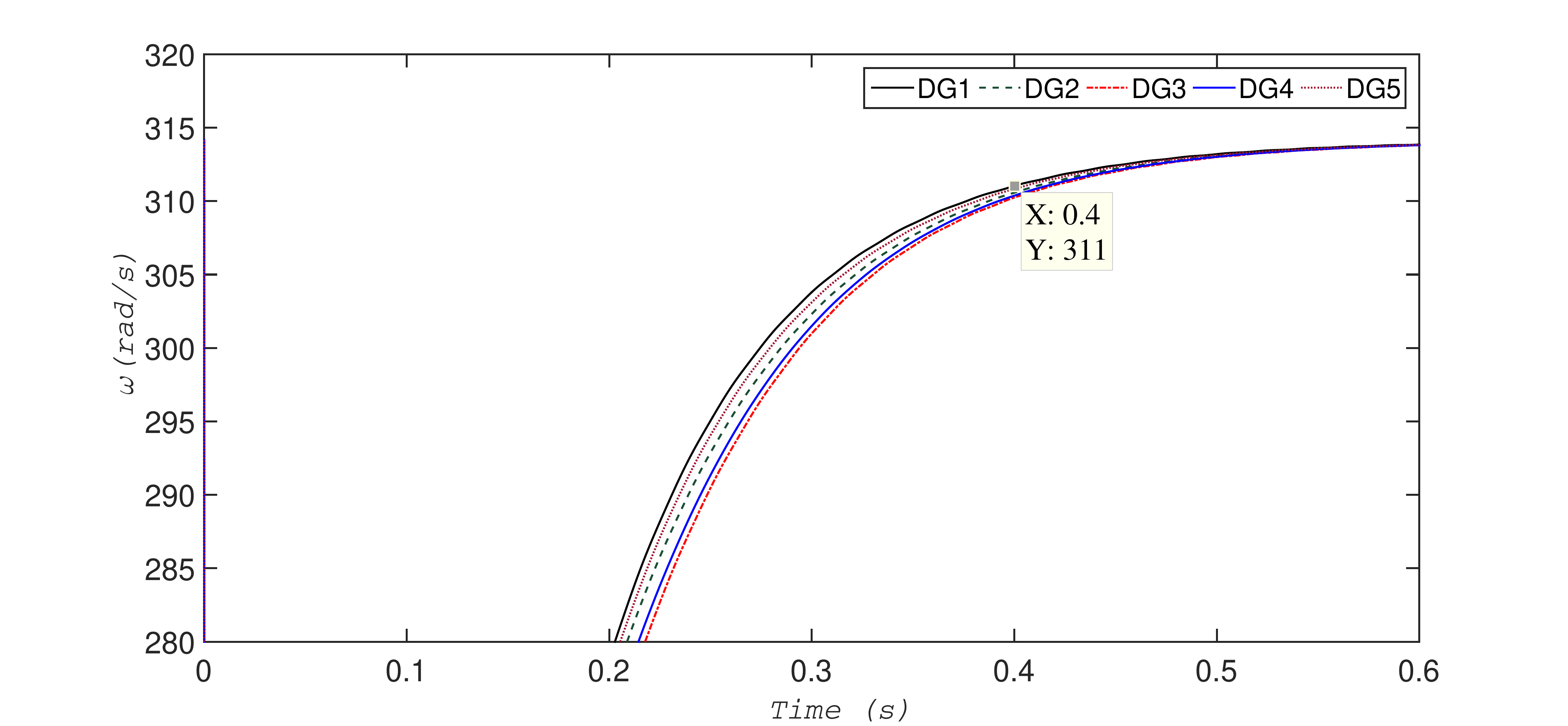} \caption{Convergence rate problem with $\protect\lambda^\star =10$: Voltage
	(top) and frequency (bottom) evolution of 5-bus system. Algorithm 2
	results in pinning DG3.}
\label{fig:ConvRate1}\vspace{+10pt} %
\includegraphics[width=8.5cm,height=3.8cm]{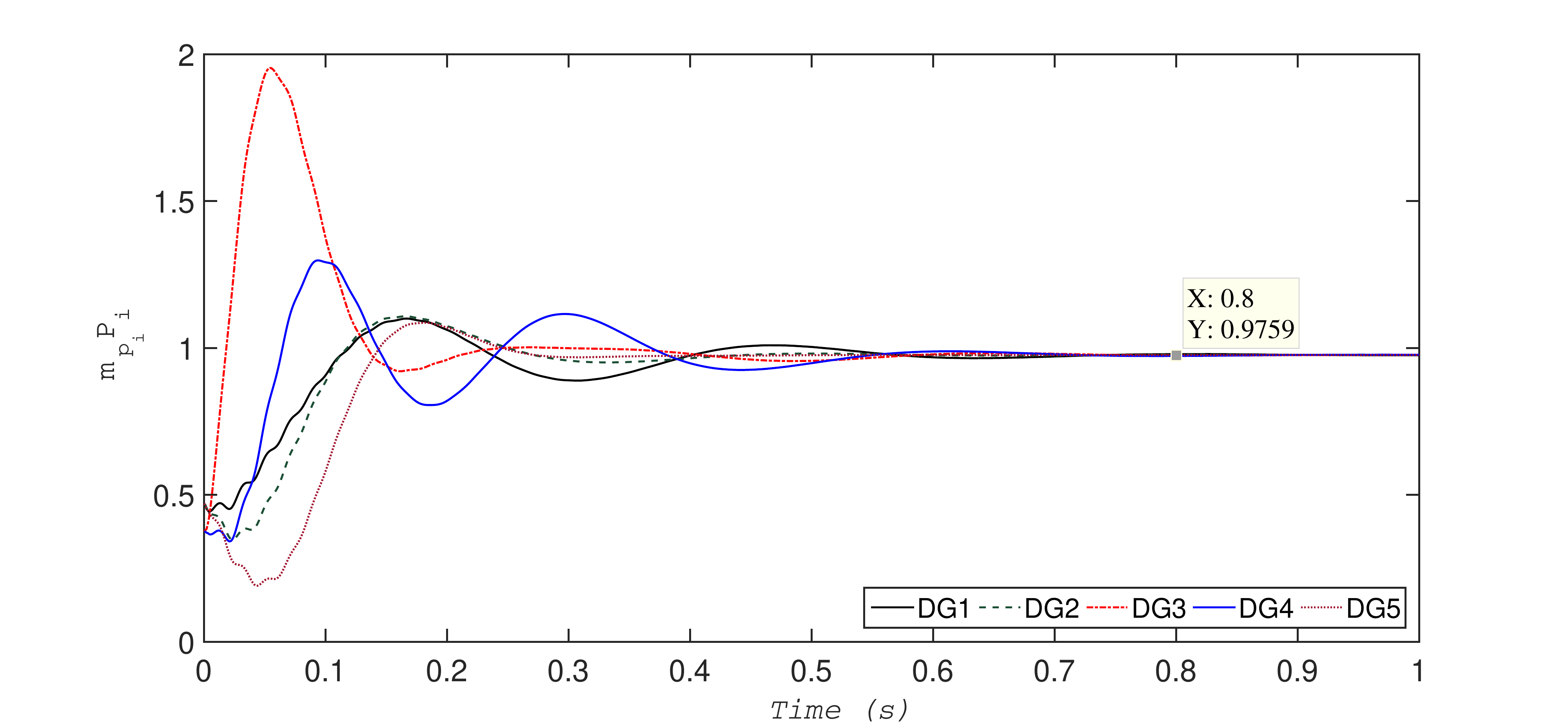} 
\caption{DGs output power for pinning DG3 in power system given in Fig. 
\protect\ref{fig:5b}. }\vspace{-10pt}
\label{fig:5ringPower}
\end{figure}

For our numerical simulations, we have used the Simpower System Toolbox of
Simulink
for 4 bus and 5 bus power systems, shown in Figs. \ref{fig:4buspinning} and %
\ref{fig:5busRing}, with different topologies and communication networks to
show the adaptability and effectiveness of the proposed pinning control
method. Microgrid operates on a 3-phase, 380V(L-L) and frequency of 50 Hz ($%
\omega _{0}=314.15(rad/s)$). Unequal parameters of the DGs are given in the
top section of Table \ref{tab: DGs} while the remaining parameters are
adopted from \cite{Poga}. In all case studies, DG 3 and DG 4 are assumed to
be Type II DGs and the rest of the DGs are considered to be Type I. Loads
are as given in the bottom section of Table \ref{tab: DGs}.

As mentioned earlier, an ideal DC source is assumed from DG side, therefore,
the weather effect is not considered in this study. The cut-off frequency of
the low-pass filters, $\omega _{c}$, is set to $31.41(rad/s)$. The control
gains $c_{v}$, $c_{\omega }$ and $c_{p}$ are all set to $400$. It should be
noted that microgrid islanding operation is detected based on the status of
the main breaker and disconnect switches at utility/grid point of
connection. We note here that the undershoot/overshoot of voltage amplitude
and frequency of the DGs in microgrid during the transient from grid
connected to islanding mode should not exceed 10-20 cycles to avoid the
operation of 27, 59, and 81 protective relays. Generally, the protective
power relays for voltage and frequency are typically set to $0.88(p.u.)\leq
v_{\text{mag}}\leq 1.1(p.u.)$ and $295.3(rad/s)\leq \omega \leq 317.3(rad/s)$
for 10-20 cycles. The microgrid's main breaker opens at $t=0(s)$ and goes to
islanding mode at which time the secondary voltage and frequency control are
activated.

As shown in Figs. \ref{fig:4buspinning} and \ref{fig:5busRing}, our study
cases will illustrate both two-way (undirected) and one-way (directed)
communication links in the microgrid. In one-way communication links, we
restrict the transition function so that the new state of the sender does
not depend on the current state of the receiver (a neighboring DG). Security
of the location of a DG and the criticality of the load that it feeds are
factors to be considered when deciding on directed vs undirected
communication links.

\subsection{Case 1: Pinning to Ensure Prescribed Convergence Rate}

\begin{figure}[tbp]
\centering
\includegraphics[width=8.5cm,height=3.8cm]{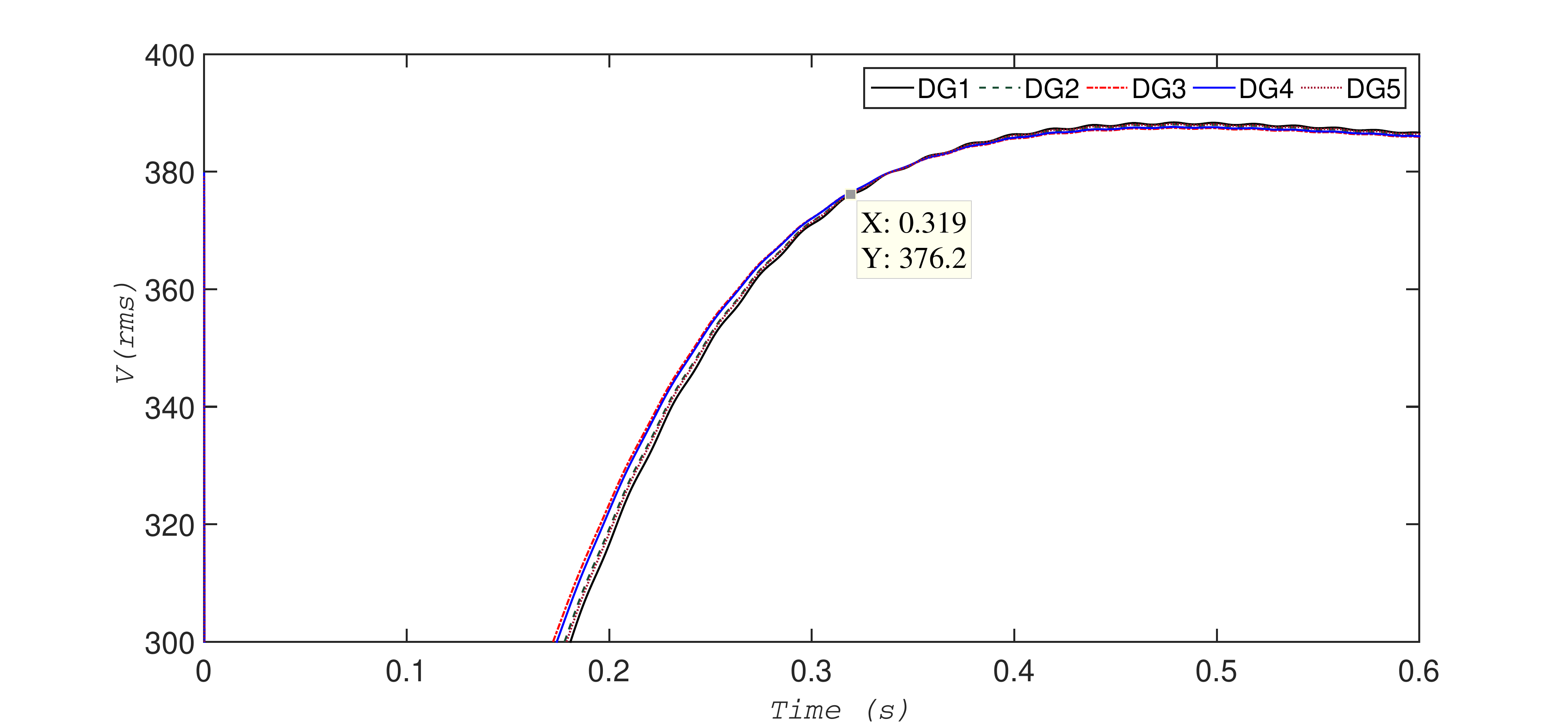} %
\includegraphics[width=8.5cm,height=3.8cm]{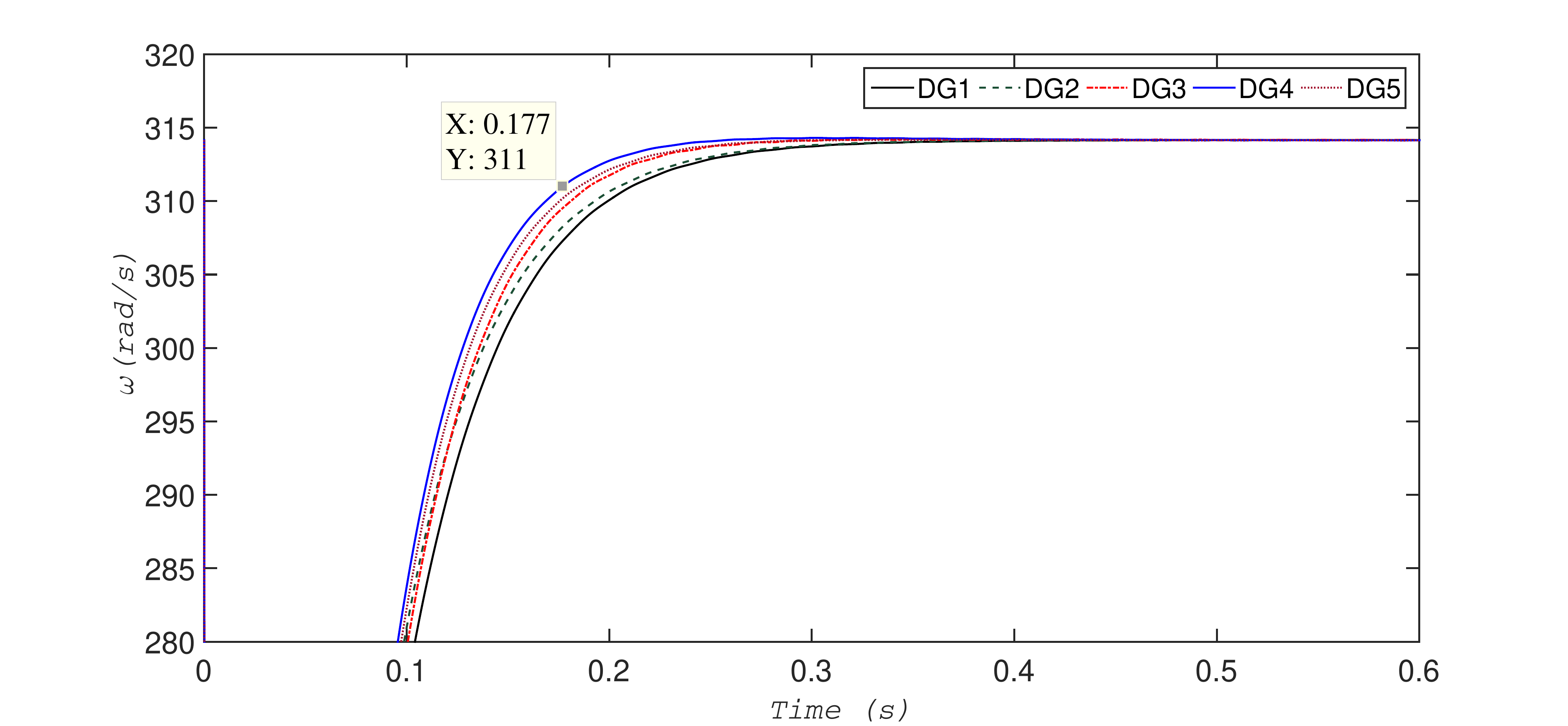}
\caption{Convergence rate problem with $\protect\lambda^\star =20$: Voltage
(top) and frequency (bottom) evolution of 5-bus system. Algorithm 2
results in simultaneous pinning of DG1 and DG3.}
\label{fig:ConvRate2}
\end{figure}

\begin{table}[t]
\centering
\begin{tabular}{l}
\begin{tabular}{|p{1.2cm}|p{2.2cm}|p{2.2cm}|}
\hline\hline
& \vspace{1pt}Type I & \vspace{1pt} Type II \\[5pt] \hline
\vspace{1pt}$m_P$ & \vspace{1pt}$9.4\times10^{-5}$ & \vspace{1pt} $12.5
\times10^{-5}$ \\[5pt] 
$n_Q$ & $1.3\times10^{-3}$ & $1.5 \times10^{-3}$ \\[5pt] \hline\hline
\end{tabular}
\\ 
\begin{tabular}{|p{3cm}|p{3cm}|}
\vspace{2pt}Load 1 & \vspace{2pt} Load 2 \\[5pt] \hline
\vspace{2pt} $12KW +j 12 KVar$ & \vspace{2pt}$15.3KW + j 7.6KVar$ \\%
[5pt] \hline\hline
\end{tabular}%
\end{tabular}%
\caption{DG and Load parameters of the systems.}
\label{tab: DGs}
\end{table}
Let the pinning gain be $g=0.2$ and the desired convergence rate be $\lambda
^{\star }=10$. From the formulation of Problem 2, we can find the desired
algebraic connectivity to the reference to be $\mu ^{\star }\geq \lambda
^{\star }/c_{v}=0.025$ for voltage\footnote{%
Since $c_{v}$ and $c_{\omega }$ are chosen the same, this calculation holds
for frequency as well.}. Running Algorithm 2 gives two single pinning
solutions, \textit{viz.,} $\mathcal{P}=\{\text{DG2}\}$ or $\mathcal{P}=\{%
\text{DG3}\}$; in the first iteration of step 4, we have $\text{path}({DG1}%
,\,\mathcal{N}\setminus \{DG1\})=\text{path}({DG4},\,\mathcal{N}\setminus
\{DG4\})=7$, $\text{path}({DG2},\,\mathcal{N}\setminus \{DG2\})=\text{path}({%
DG3},\,\mathcal{N}\setminus \{DG3\})=6$, and $\text{path}({DG5},\,\mathcal{N}%
\setminus \{DG5\})=\infty $. Since out degree of both DG2 and DG3 are the
same, the algorithm predicts that the performance for pinning either one of
these DGs should be identical. As determined by the algorithm, pinning DG5
will not help microgrid synchronization because it does not share any
information with any of its neighboring DGs. Fig. \ref{fig:ConvRate1} shows
the results for this case. The exponential convergence rate can be
calculated as $\lambda _{\omega }=\ln (0.01)/t_{ss_{f}}=4.6/0.4=11.5$ and $%
\lambda _{V}=\ln (0.01)/t_{ss_{V}}=4.6/0.319=14.37$ for frequency and
voltage, respectively. The output powers of the DGs for pinning of DG3 are
shown in Fig. \ref{fig:5ringPower} to illustrate proper power sharing during
islanded operation of the microgrid.

Now, if the desired convergence rate is set to $\lambda ^{\star }=20$, we
find the required algebraic connectivity to be $\mu ^{\star }=0.05$.
Algorithm 2 results in a solution pinning two DGs ($m=2$), \textit{viz.,} $%
\mathcal{P}=$ $\{\text{DG1, DG3}\}$: in the first iteration, as before,
either DG2 or DG3 should be selected. If DG2 is selected at the start of the
second iteration, we have $\mathcal{I}=\{\text{ DG1, DG3, DG4, DG5}\}$ and $%
\text{path}(\{\text{DG2, DG1}\},\,\mathcal{I}\setminus \{\text{ DG1}\})=%
\text{path}(\text{DG2, DG3},\,\mathcal{I}\setminus \{\text{DG3}\})=\text{path%
}(\{\text{DG2, DG5}\},\,\mathcal{I}\setminus \{\text{ DG5}\})=4$ and $\text{%
path}(\{\text{DG2, DG4}\},\,\mathcal{I}\setminus \{\text{DG4}\})=3$.
Therefore, for $m=2$ the pinning is $\mathcal{P}=\{\text{DG2, DG4}\}$. If in
the first iteration, DG3 is selected, then $\text{path}(\{\text{DG2, DG4}%
\},\,\mathcal{I}\setminus \{\text{DG1}\})=3$ and the pinning set $\mathcal{P}%
=\{\text{DG1, DG3}\}$, which also predicts that the performance of pinning $%
\{\text{DG1, DG3}\}$ is the same as pinning $\{\text{DG2, DG4}\}$. This can
be deduced from the symmetry in the network. The results of numerical
simulations with this choice are given in Fig. \ref{fig:ConvRate2}. The
respective convergence rates for voltage and frequency in this case can be
computed to be $26.7$ and $26$, respectively.

\begin{figure}[!t]
\centering
\includegraphics[width=8.5cm,height=3.8cm]{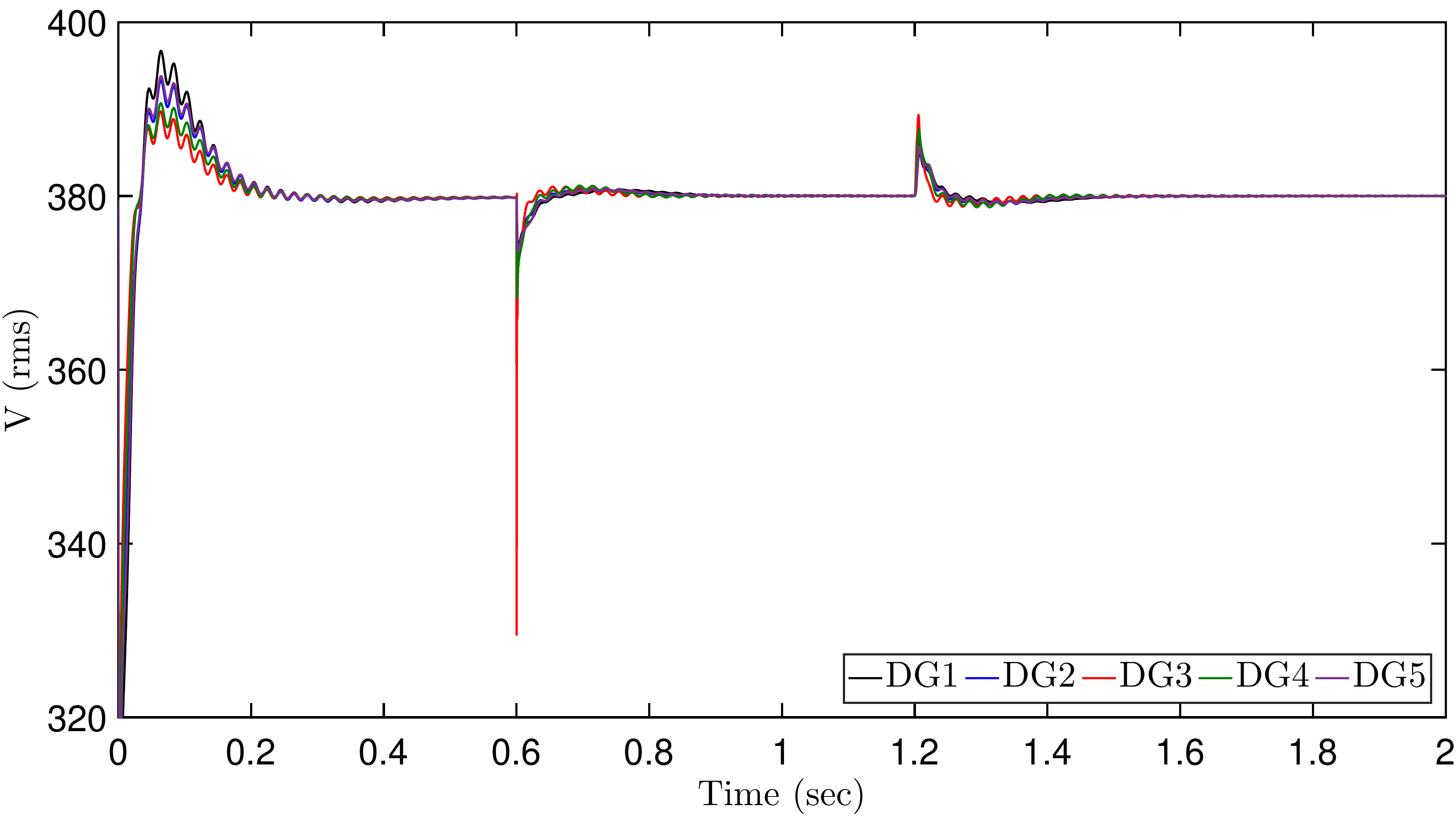} %
\includegraphics[width=8.9cm,height=3.8cm]{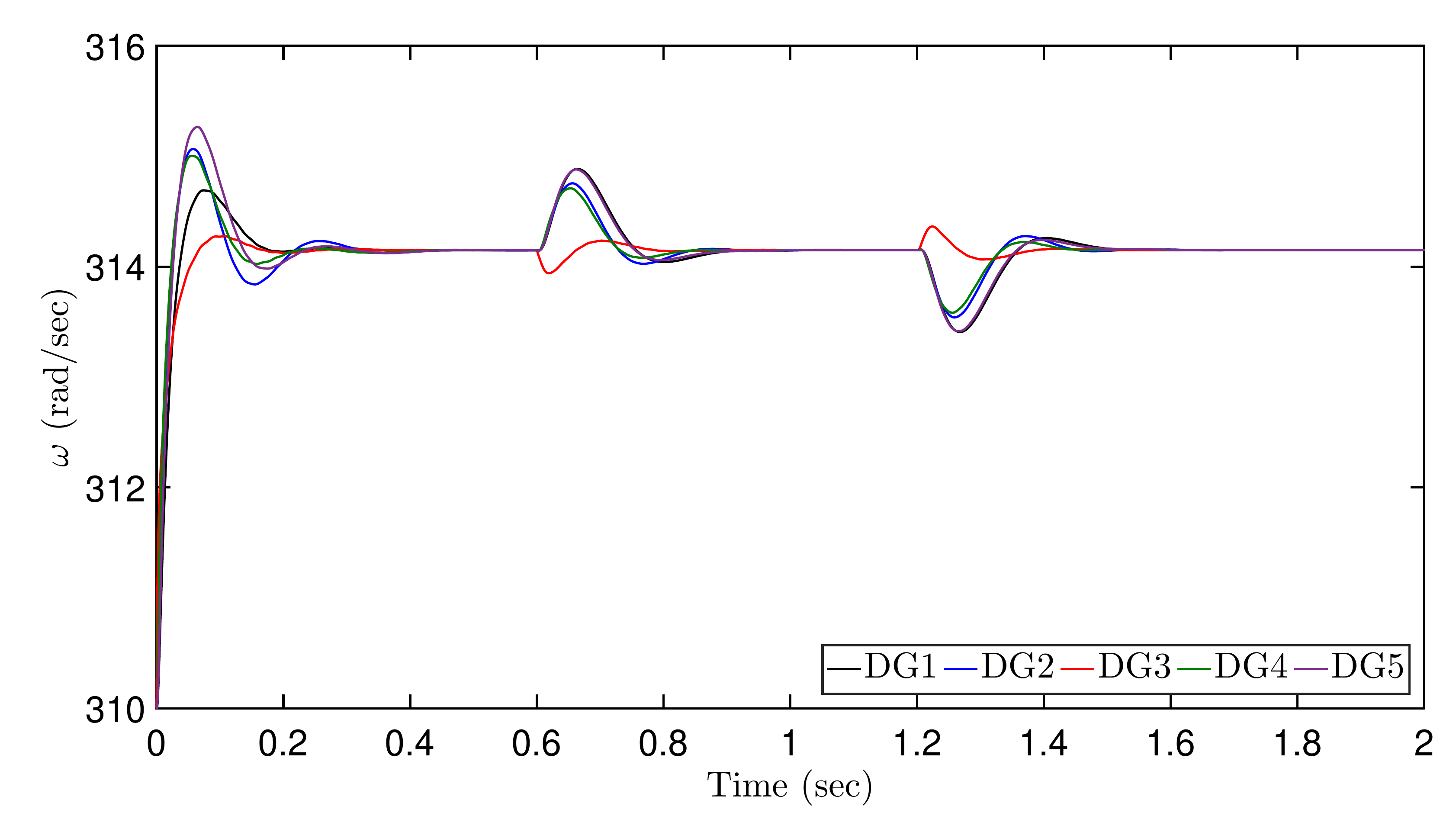}
\caption{Load Variation Scenario: voltage and frequency evolution under pinning DG3 (proposed for single pinning) with load increment and decrement.}
\label{fig: OL}
\end{figure}

\subsection{Case 2: Load Variation}

\begin{figure}[t]
\includegraphics[width=8.5cm,height=3.8cm]{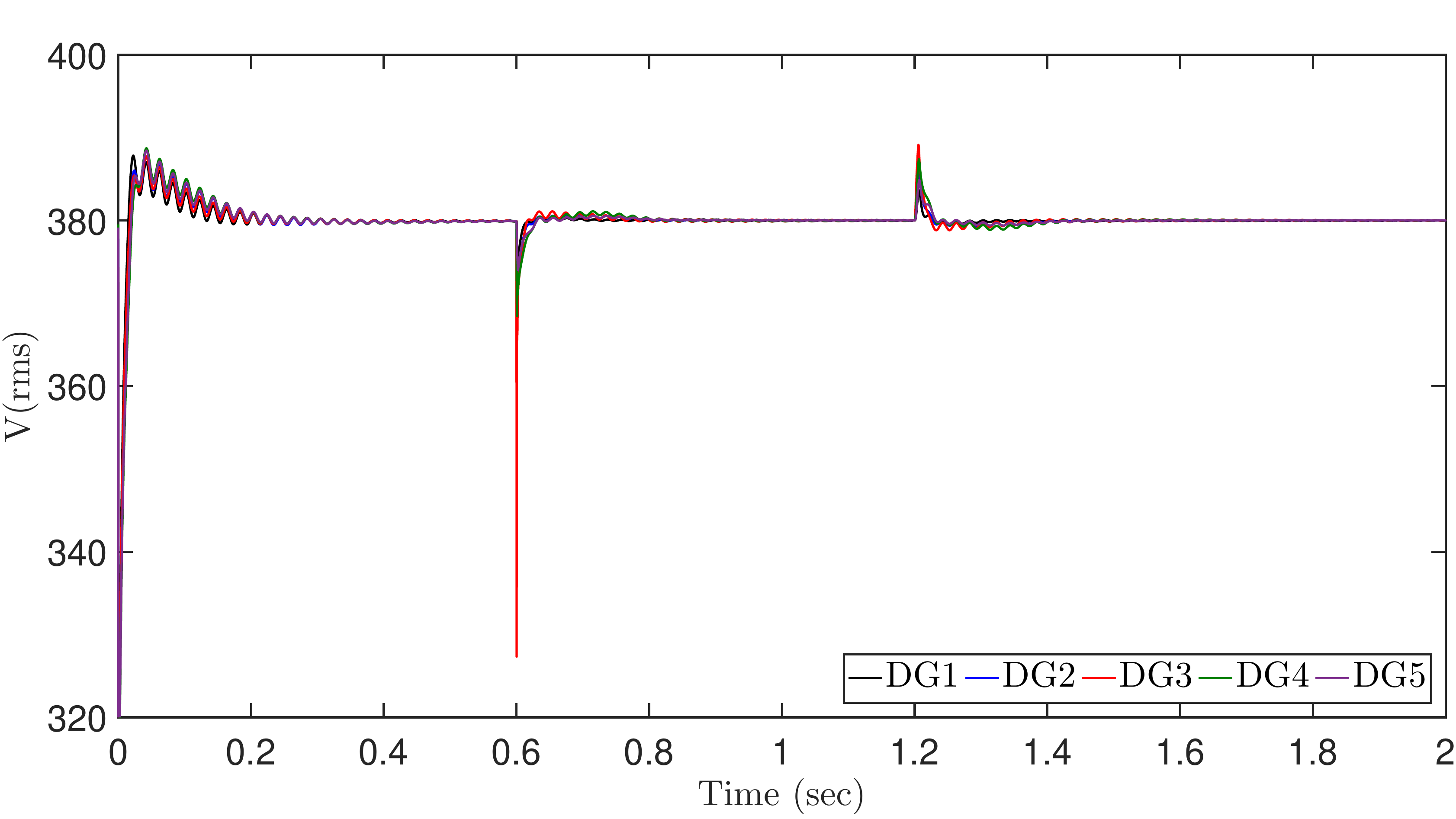} %
\includegraphics[width=8.5cm,height=3.8cm]{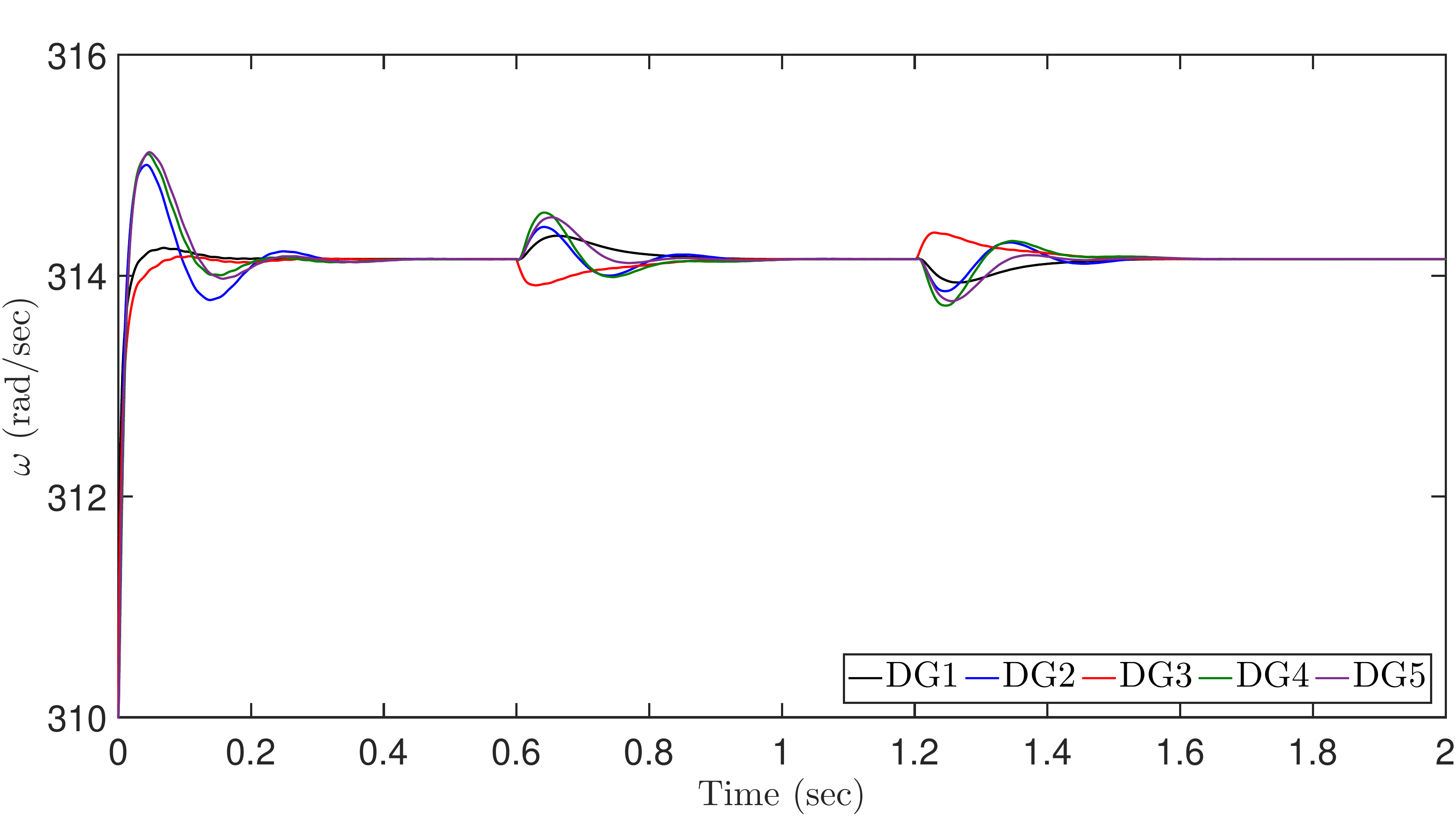} 
\caption{Load Variation Scenario: voltage and frequency evolution under pinning DG1 and DG3 (proposed for multiple pinning) with load increment and decrement.}
\label{fig:loadTr}
\end{figure}

In Fig. \ref{fig:loadTr}, a severe scenario for load variation of the
microgrid in Fig. \ref{fig:5busRing} is considered to show the effectiveness
of our proposed intelligent single and multiple pinning algorithms. As was
mentioned earlier, our proposed algorithm chose $\mathcal{P}=\{$DG3$\}$ for
a single pinning solution and selected pinning sets $\mathcal{P}=\{\text{
DG1, DG3}\}$ or $\mathcal{P}=\{\text{DG2, DG4}\}$) for a multiple pinning
solution. In the load variation scenario, the primary load on Bus 3 is fixed
with power factor of $P.F.=0.85$; then, another mostly active load of $%
12\,KW\,+\,j1\,KVar$ is added at 0.6 $(sec)$ and removed at $1.2~(sec)$. As
can be observed from the results shown in Figs. \ref{fig: OL} and \ref%
{fig:loadTr}, our intelligent pinning algorithm based solutions are
comparatively more robust under this load variation scenario and bring back
the voltage and frequency of the microgrid to the reference value very fast.
Results also indicate the superior performance of the multiple pinning
algorithm in comparison with single pinning algorithm during load variation.

\subsection{Case 3: Comparative Study}

Here, we assume that the network is to be stabilized by single pinning
method. The bus and communications networks are given in Fig. \ref%
{fig:4buspinning}. In this configuration, it is assumed that the DGs
communicate with each other through a fixed communication network shown in
Fig.~\ref{fig: 4busCom}. The diagram shows that the DGs only communicate
with their neighboring DG. The DGs' terminal voltage amplitude and frequency
for different reference single pinning scenario are shown in Fig. \ref%
{fig:4busVT}. Based on the tracking synchronization control strategy, it can
be seen that all DGs' terminal voltage and frequency return to the reference
value dictated by the leader DG. However, pinning DG2 results in a faster
and more robust convergence in comparison with DG1 presented in \cite{Bidram}%
. Please note that in \cite{Bidram}, because of its minimum directed
communication topology, pinning DG1 is suggested while our pinning algorithm
indicates that DG2 should be pinned which also coincides with the optimal
solution of Problem 1.

\begin{figure*}[t]
\centering
\begin{subfigure}{1\textwidth}
		\centering
		\includegraphics[width=8.5cm,height=3.8cm]{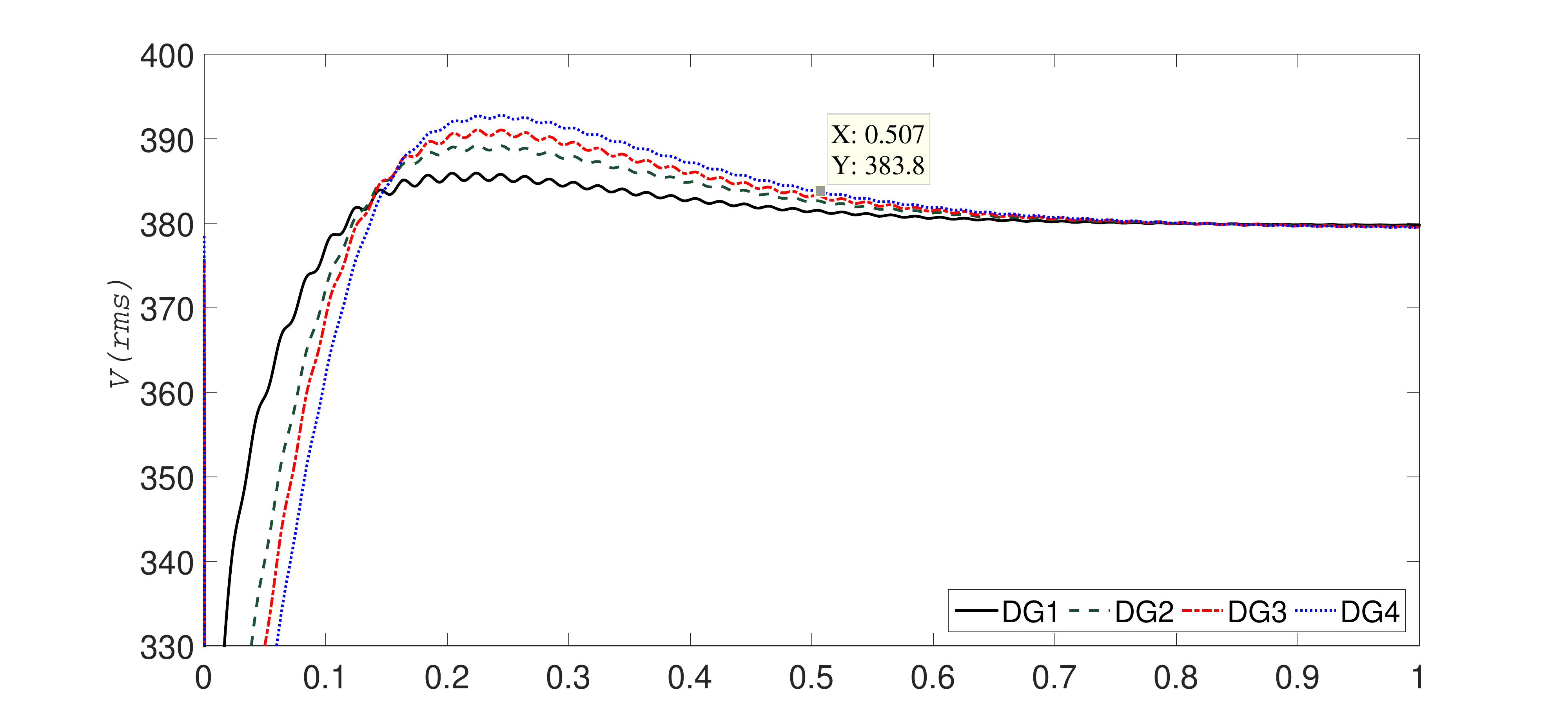}
		\includegraphics[width=8.5cm,height=3.8cm]{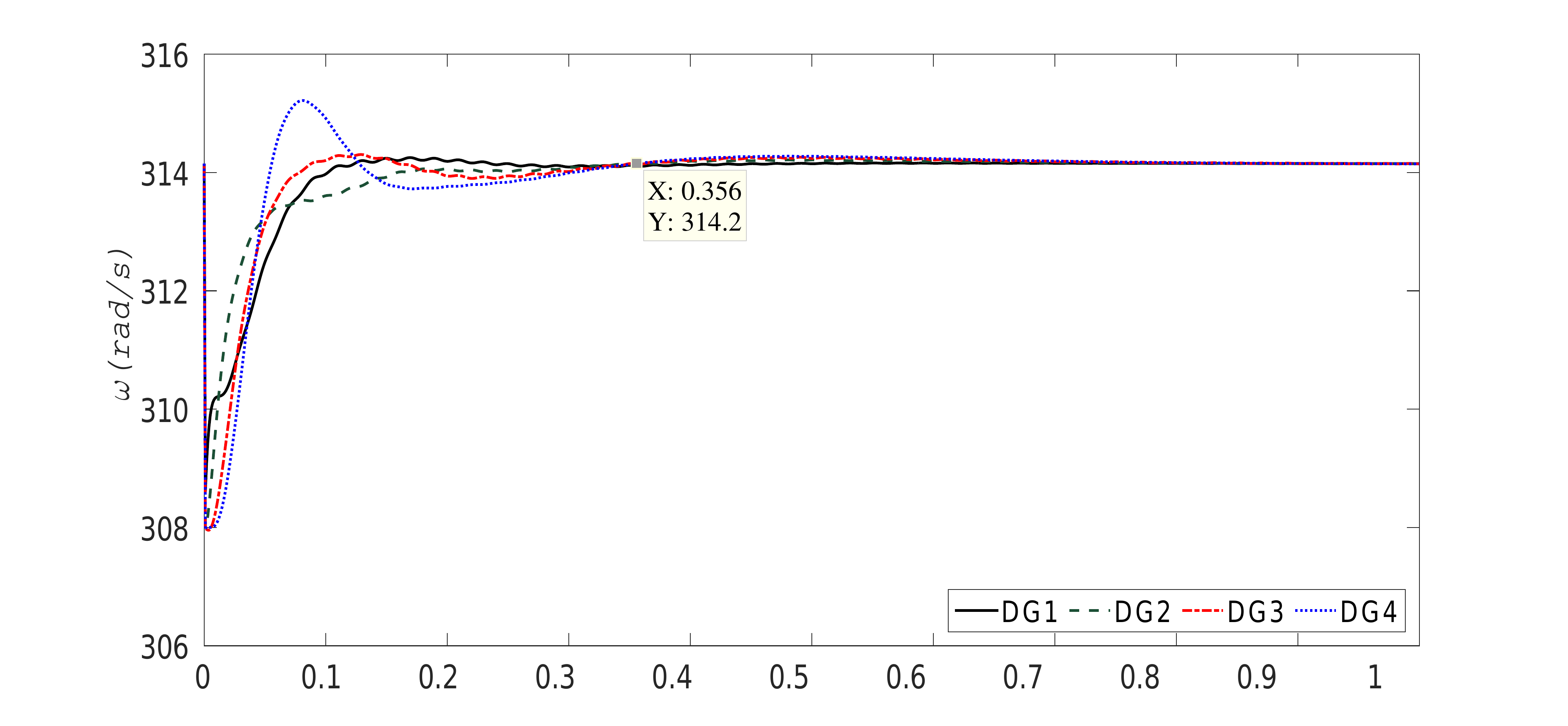}
		\caption{pinning DG1}
		\label{}
	\end{subfigure} \hspace{0.05\textwidth} 
\begin{subfigure}{1\textwidth}
		\centering
		\includegraphics[width=8.5cm,height=3.8cm]{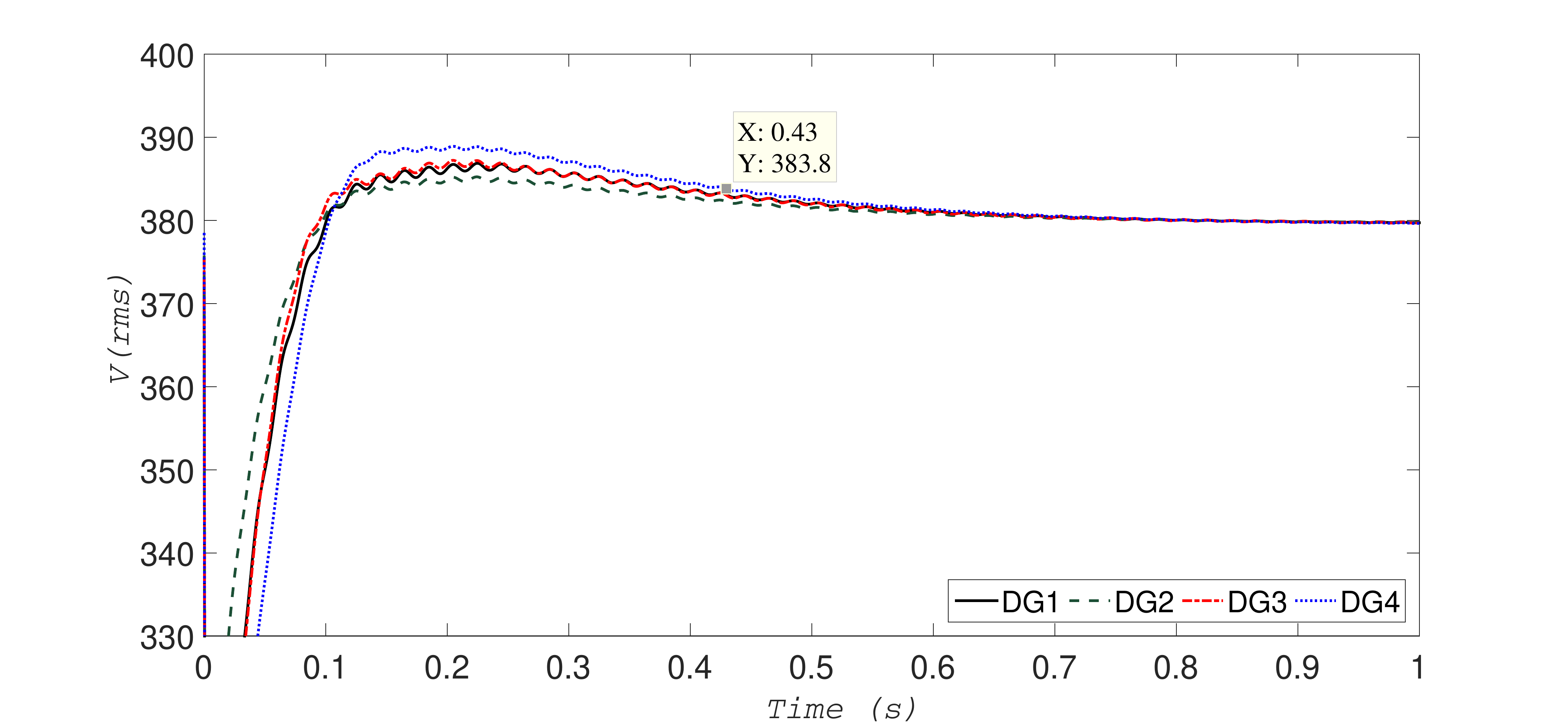}
		\includegraphics[width=8.5cm,height=3.8cm]{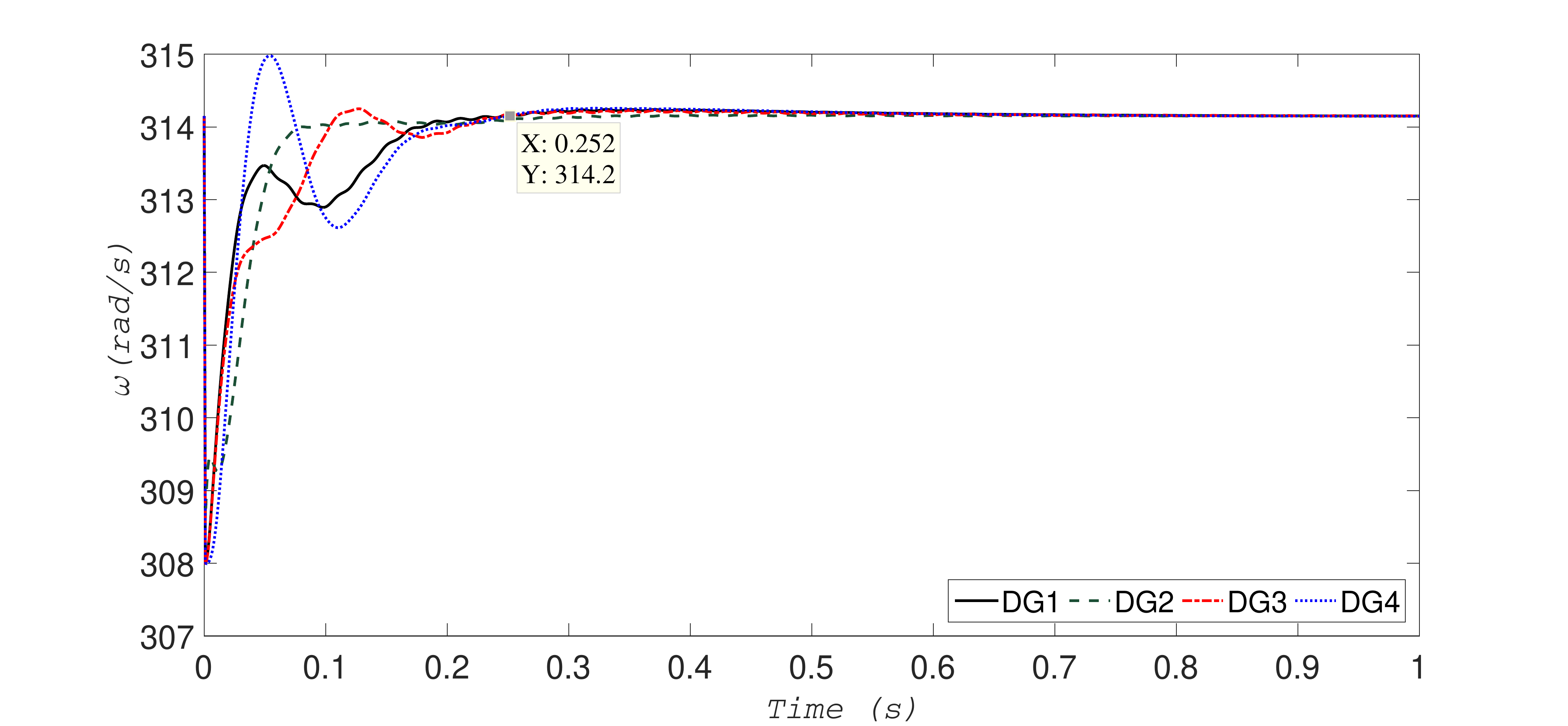}
		\caption{pinning DG2 (proposed)}
		\label{fig:4busDG2}
	\end{subfigure} 
\caption{DGs terminal amplitudes voltages (at left) and frequency (at right)
for two pinning selections in 4-bus configuration corresponding to Fig. \protect\ref{fig:
4busCom}.}
\label{fig:4busVT}
\end{figure*}

Table \ref{tab:4bus} provides information about settling time of DGs'
terminal voltage and frequency for different pinning cases. As can be
observed in Fig. \ref{fig:4busDG2}, pinning DG2 results in better transient
behavior, compared to pinning the other DGs in the network. It should be
noted that DG4 cannot be selected as a leader because it does not share
information with the rest of the microgrid. This is also true for pinning
DG5 in 5-bus power systems in Fig. \ref{fig:5busRing}.

\begin{table}[t]
\center
\begin{tabular}{|p{2cm}|p{1.12cm}|p{1.12cm}|p{1.2cm}|}
\hline
Pinning DG & ${t_{s_v}}~(s)$ & ${t_{s_f}}~(s)$ & $\text{ path}(\mathcal{P}, 
\mathcal{I})$ \\ \hline
DG1 & 0.51 & 0.36 & 6 \\ 
DG2 & 0.43 & 0.25 & 4 \\ \hline
\end{tabular}%
\caption{Single pinning of 4 bus system given in Fig. \protect\ref%
{fig:4buspinning}.}
\label{tab:4bus}
\end{table}

\section{Conclusions}

Intelligent single and multiple pinning based distributed cooperative
control algorithms have been proposed to efficiently synchronize DGs in a
microgrid to their nominal voltage and frequency values after disconnecting
from the main grid. It has been shown that selection of pinning nodes
depends directly on the power system and communication network topologies.
Case studies using different types of microgrid configurations and scenarios
demonstrate that the proposed methodology helps the DGs in the microgrid
achieve convergence with desired rate and improved transient voltage and
frequency behavior after going to islanding mode.

\appendices

\section{Bounds on $\protect\phi (\mathbf{Z})$\label{app: bounds}}

\begin{figure}[t]
\centering
\includegraphics[width=
	.7\columnwidth]{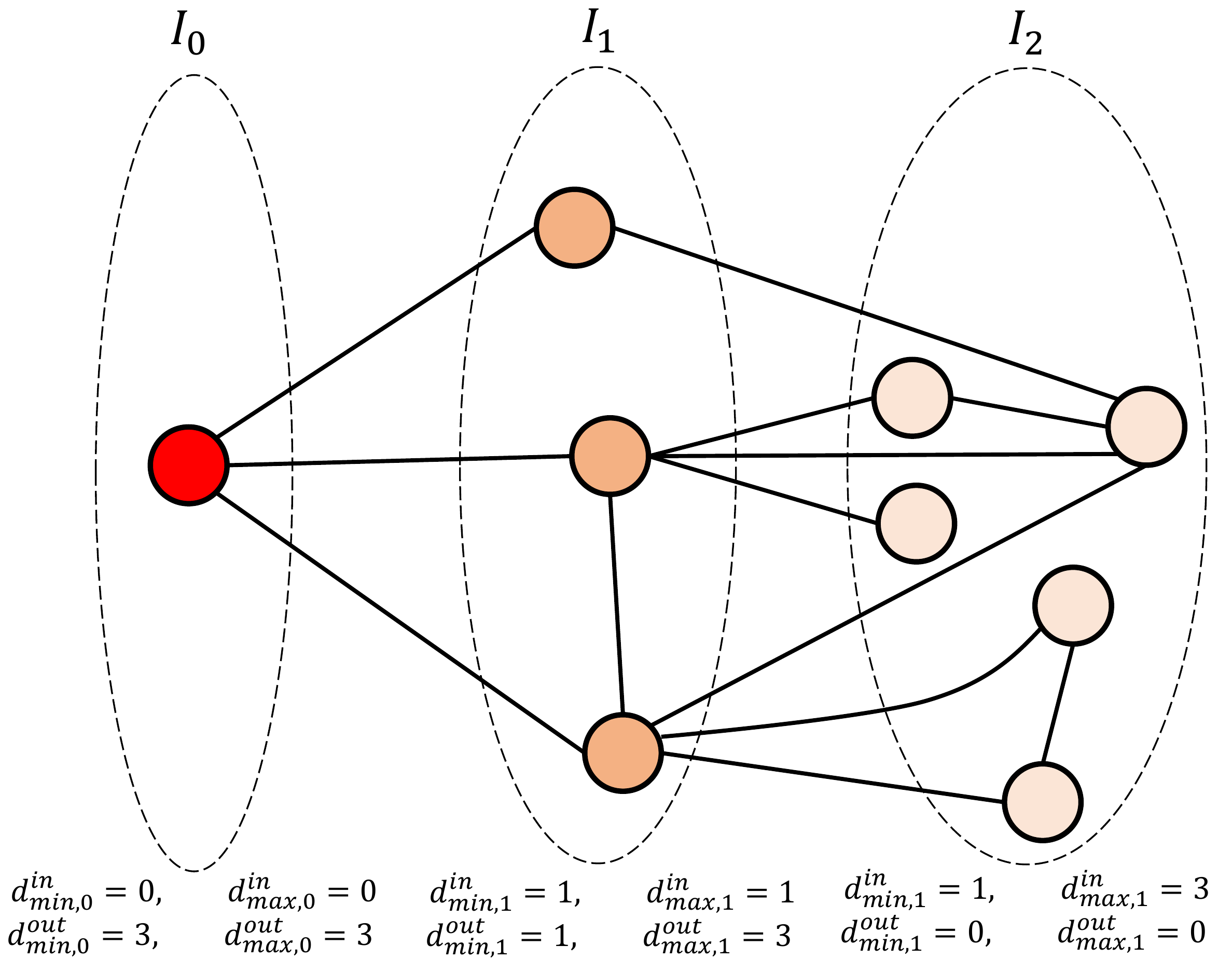} \vspace{0.4cm}
	\caption{Sample topology for single pinning where the farthest node from the
		pinning node, $\mathcal{I}_{0}$, is $k=2$. }
	\label{fig:Topology_sorting_1DG} \centering
\includegraphics[width=
	.7\columnwidth]{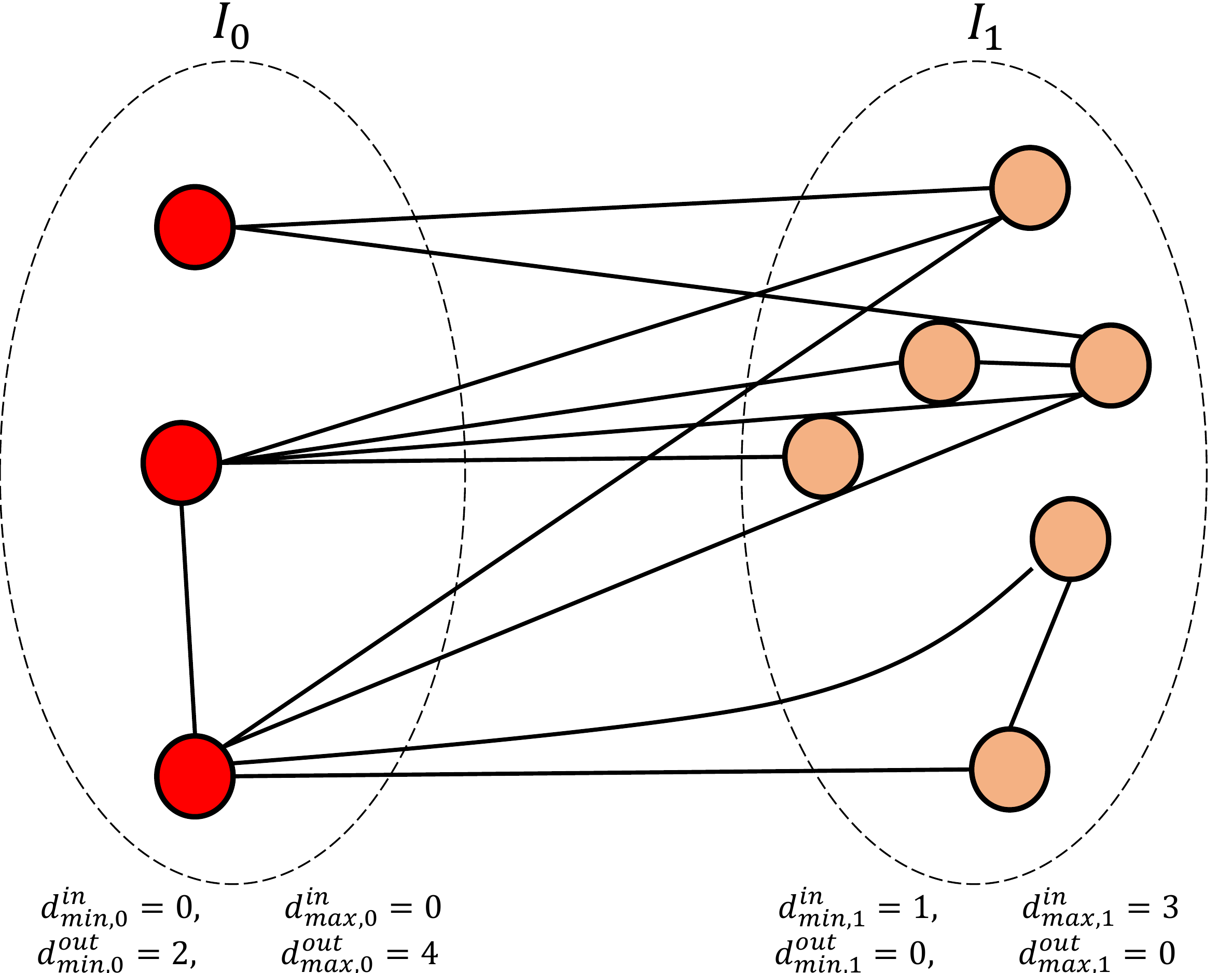}\vspace{0.4cm}
\caption{Sample topology for multiple pinning ($m=3$) where the farthest
node from the pinning set, $\mathcal{I}_{0}$, is $k=1$.}
\label{fig:Topology_sorting_3DG}
\end{figure}

Let us assume that the nodes $\mathcal{P}=\{i_{1},\,\cdots ,\,i_{m}\}$ are
pinned, this is called the pinning set. Let the farthest node to pinning
set, $\mathcal{I}_{0}$, be $k$. Then, define the set $\mathcal{I}%
_{j},\forall j\in \{1,\cdots ,k\}$ as 
\begin{equation}
\mathcal{I}_{j}=\left\{ i\Big|a_{pi}=1,\,\forall p\in \mathcal{I}%
_{j-1},\,i\in \mathcal{N}\setminus \bigcup_{r=1}^{j-1}\mathcal{I}%
_{r}\right\} ,  \label{eq: set}
\end{equation}%
where $a_{pi}$ are entries of the adjacency matrix of the communication
network, $\mathcal{N}=\{1,\,\cdots ,\,N\}$, and $\setminus $ denotes minus
operation for sets. For each set in (\ref{eq: set}), we define the following 
\begin{align}
d_{i,{j}}^{\text{out}}& =\sum_{p\in \mathcal{I}_{j+1}}a_{pi},\quad i\in 
\mathcal{I}_{j} \\
d_{i,j}^{\text{in}}& =\sum_{p\in \mathcal{I}_{j-1}}a_{ip,}\quad i\in 
\mathcal{\ I}_{j}
\end{align}%
where $\mathcal{I}_{-1}=\mathcal{I}_{k+1}=\emptyset $. Please note that due
to the definition in (\ref{eq: set}), $d_{i,j}^{\text{out}}\geq 1,\,\forall
j\in \{0,\cdots ,\,k-1\}$. From Theorem 3 in \cite[Theorem 2]{Saeed}, given
an undirected network (\textit{i.e.,} information can flow both ways) with
pinning set, $\mathcal{P}$, and $|\mathcal{P}|=m$, the algebraic
connectivity of the network $\phi (\mathbf{L},\mathbf{Z},g)$ can be bounded
as follows 
\begin{equation*}
\phi _{l}(\mathbf{L},\mathbf{Z},g)\leq \phi (\mathbf{L},\mathbf{Z},g)\leq
\phi _{u}(\mathbf{L},\mathbf{Z},g),
\end{equation*}%
where the upper bound 
\begin{equation}
\phi _{u}(\mathbf{L},\mathbf{Z},g)=\beta \left( 1-\sqrt{1-\frac{
\sum\limits_{i=1}^{m}{d_{0,\,i}^{\text{in}}}^{2}}{(N-m)\beta ^{2}}}\right)
\end{equation}%
with%
\begin{equation*}
\beta =\frac{\sum\limits_{i=1}^{m}d_{i,\,0}^{\text{out}}+(N-m)(g+d_{\min
,\,0}^{\text{in}})}{2(N-m)}
\end{equation*}%
while the lower bound, $\phi _{l}(\mathbf{L},\mathbf{Z},g)$, is the smallest
positive root of the polynomials, $\alpha _{i}(\mu )$ $i=0,\cdots ,\,k-1$ 
\begin{equation}
\alpha _{i}(\mu )=d_{\min ,\,i-1}^{\text{out}}+d_{i,\,\min }^{\text{in}}-\mu
-{d_{\max ,\,i}^{\text{in}}}^{2}/\alpha _{i+1}(\mu ),
\label{eq: lower_multiple}
\end{equation}%
where 
\begin{eqnarray*}
\alpha _{k}(\mu ) &=&d_{\min ,\,k}^{\text{in}}-\mu \\
d_{\min ,\,i}^{\text{in}} &=&\min_{j\in \mathcal{I}_{i}}(d_{j,i}^{\text{in}})
\\
d_{\min ,\,i}^{\text{out}} &=&\min_{j\in \mathcal{I}_{i}}(d_{j,i}^{\text{out}
})) \\
d_{\max ,\,i}^{\text{in}} &=&\max_{j\in \mathcal{I}_{i}}(d_{j,i}^{\text{in}})
\\
d_{\min ,\,-1}^{\text{out}} &=&g.
\end{eqnarray*}%
Figs. \ref{fig:Topology_sorting_1DG} and \ref{fig:Topology_sorting_3DG},
respectively, illustrate the sets and variables defined above for sample
single ($m=1$) and multiple ($m=3$) pinning cases. 

\end{document}